\documentclass[useAMS,usenatbib]{mn2e}
\pdfoutput=1
\usepackage{graphicx}
\usepackage{dcolumn}
\usepackage{bm}
\usepackage{amssymb,amsmath,bm}
\usepackage{color}
\usepackage[colorlinks,linkcolor=red,citecolor=blue,urlcolor=blue ]{hyperref}
\usepackage{multirow}
\usepackage[utf8]{inputenc}
\usepackage{balance}
\usepackage{enumitem}
\usepackage{lipsum}
\usepackage{ulem}

\title[Investigating the nonthermal emission of eHWC J2019$+$368]{Investigating the multiband nonthermal emission of the 100 TeV  source eHWC J2019$+$368 with a pulsar wind nebula scenario}
\author[Jun Fang et al.]{Jun Fang$^{1,3}$,  Lu, Wen$^1$, Huan Yu$^2$, Chen Songzhan$^3$\\
                      $^{1}$Department of Astronomy, Yunnan University, Kunming 650091, China; fangjun@ynu.edu.cn\\
                      $^{2}$Department of Physical Science and Technology, Kunming University, Kunming 650214, China; yuhuan.0723@163.com  \\
                      $^{3}$Institute of High Energy Physics, Chinese Academy of Sciences, Beijing 100049, China
                      }

\begin{document}
  \date{\today}
  \pagerange{1--6} \pubyear{2020}
  \maketitle

\begin{abstract}
  eHWC J2019+368 is one of the sources emitting $\gamma$-rays with energies higher than 100 TeV based on the recent measurement with the High Altitude Water Cherenkov Observatory (HAWC), and the origin is still in debate. The pulsar PSR J2021$+$3651  is spatially coincident with the TeV source. We investigate theoretically whether the multiband nonthermal emission of eHWC J2019+368 can originate from the pulsar wind nebula (PWN) G75.2$+$0.1 powered by PSR J2021$+$3651. In the model, the spin-down power of the pulsar is transferred to high-energy particles and magnetic field in the nebula. As the particles with an energy distribution of either a broken power-law or a power-law continually injected into the nebula, the multiband nonthermal emission is produced via synchrotron radiation and inverse Compton scattering. The spectral energy distribution of the nebula from the model with the reasonable parameters is generally consistent with the detected radio, X-ray and TeV $\gamma$-ray fluxes. Our study supports that the PWN has the ability to produce the TeV $\gamma$-rays of eHWC J2019+368, and the most energetic particles in the nebula have energies up to about $0.4$ PeV.
\end{abstract}

\begin{keywords}
  ISM: individual objects (G75.2$+$0.1) -- radiation mechanisms: non-thermal
\end{keywords}

\section{Introduction}\label{sec:intro}
A source emitting $\gamma$-rays with energies above 100 TeV is a potential candidate which can accelerate cosmic-rays to the "knee" at $\sim 1 $\,PeV. Detecting and identifying them can shed light on the primarily contributor to the Galactic cosmic-rays. Recently, \citet{2020prl...124..021102A} reported that nine sources had been detected above 56 TeV with HAWC, and three of them, eHWC J1825-134, eHWC J1907+063 and eHWC J2019+368, emit $\gamma$-rays past 100 TeV.

The 100 TeV source eHWC J2019+368 has an integral flux of $\sim 1.6\times 10^{-14}\, \mathrm{ph}\,\mathrm{cm}^{-2}\,\mathrm{s}^{-1}$ above 56 TeV based on the HAWC results \citep{2020prl...124..021102A}. It is within the Cygnus region, which is composed of some potential cosmic-ray accelerators \citep{2007ApJ...658L..33A,2009ApJ...695..883B,2012ApJ...745L..22B,2018ApJ...861..134A}. This source was firstly detected with Milagro (MGRO J2019$+$37) in TeV $\gamma$-rays with a flux of $\sim 80\%$ of the Crab nebula \citep{2012ApJ...753..159A}. Moreover, the observation with VERITAS indicated an extended source VER J2016$+$378 constitutes the bulk of the TeV emission, and the $\gamma$-ray region contains the energetic pulsar PSR J2021$+$3651 and the star formation region Sh 2-104 \citep{2014ApJ...788...78A}.

GeV $\gamma$-rays from PSR J2021$+$3651 have been detected with {\it Fermi}-LAT, and the upper limit on the GeV $\gamma$-rays from the putative PWN named G75.2$+$0.1 \citep{2004ApJ...612..389H,2008ApJ...680.1417V},  is $<10\%$ of the pulsed emission \citep{2009ApJ...700..1059A}.  \citet{2017ApJ...841..104M} performed X-ray studies on VER J2016$+$378 using the data of {\it Suzaku} and {\it XMM-Newton}. The X-ray nebula was indicated to have an extension of more than $27'$ along the major axis, and the association of it with the TeV source VER J2016$+$378 was supported. Based on the multi-band data obtained with {\it XMM-Newton}, {\it INTEGRAL}, and {\it Fermi}-LAT, \citet{2014ChPhc.38.085001H} proposed that the $\gamma$-rays of MGRO J2019$+$37 might be produced by the high-energy electrons in G75.2$+$0.1.

Terminate shocks in PWNe are generated due to the interaction of the relativistic winds blown from the pulsars with the ambient medium, and the energy of the flow can be efficiently transferred into electrons/positrons and magnetic field \citep{2006araa.44.17G,2020arXiv200104442A}.  Particle in cell simulations on relativistic shocks indicated that the electrons/positrons could be accelerated to high-energies when they bounce around the shock \citep{2008ApJ...682L...5S,2019MNRAS.485.5105C}. Studies on PWNe indicated that they had the ability to produce multi-wavelength nonthermal emission from radio to TeV $\gamma$-rays via synchrotron radiation and inverse Compton scattering.

In this paper, we investigate whether the nebula G75.2$+$0.1 powered by the pulsar PSR J2021$+$3651 has the ability to produce the TeV flux of eHWC J2019+368 with appropriate parameters based on a time-dependent model for the multiband nonthermal emission from PWNe. In Section \ref{sect:model}, the detail of the model is described. The results on the nebula are presented in Section \ref{sect:results}. Finally, the discussion and summary are given in Section \ref{sect:discon}.

\section{The model for the multiband nonthermal emission from PWNe}
\label{sect:model}

In a PWN, high-energy leptons (electrons and positrons) are continually injected when the pulsar spins down gradually. The lepton distribution $N(\gamma, t)$ at time $t$ can be solved from the diffusion equation \citep[e.g.,][]{2010AA...515A..20F,2012mnras.427.415m},
\begin{equation}
\frac{\partial N(\gamma, t)}{\partial t} = \frac{\partial}{\partial\gamma}[\dot{\gamma}(\gamma, t)N(\gamma, t)] - \frac{N(\gamma, t)}{\tau(\gamma, t)} + Q(\gamma, t) \;,
\label{eq:ngammat}
\end{equation}
where $\dot{\gamma}(\gamma, t)$ is the energy loss rate of the lorentz factor of the leptons, $\tau(\gamma, t)$ is the escape time due to Bohm diffusion \citep{2008ApJ...676..1210Z}, and $Q(\gamma, t)$ represents the injection of the particles per unit time in the whole nebula.

For a pulsar with period $P$, period-derivative $\dot{P}$, breaking index $n$ and age $t_{\mathrm{age}}$, the characteristic age is $\tau_c = P/2\dot{P}$, and the spin-down luminosity is
\begin{equation}
L(t) = 4\pi^2 I \frac{P}{\dot{P}^3},
\label{eq:lt}
\end{equation}
where the moment of inertia of the pulsar $I$ is adopted to be $10^{45}\mathrm{g}\,\mathrm{cm}^{2}$ in this paper.
The injection luminosity equals $L(t)$, and it follows \citep{2006araa.44.17G}
\begin{equation}
L(t) = L_0 \left ( 1+\frac{t}{\tau_0}\right )^{-\frac{n+1}{n-1}},
\label{eq:ltij}
\end{equation}
where $L_0$ is the initial luminosity, and the initial spin-down time scale of the pulsar is
\begin{equation}
\tau_0 = \frac{2\tau_c}{n-1} - t_{\mathrm{age}}.
\label{eq:tau0}
\end{equation}

The spin-down luminosity is transferred into particles and magnetic field. Assuming the magnetic energy fraction is $\eta$, the magnetic field strength in the PWN is \citep{2010apj.715.1248T,2012mnras.427.415m}
\begin{equation}
B(t) = \frac{3(n-1)\eta L_0 \tau_0}{R^3_{\mathrm{PWN}}(t)} \left [1-\left( 1+\frac{t}{\tau_0}\right )^{\frac{2}{n-1}}\right ],
\label{eq:bt}
\end{equation}
and the radius of the nebula in the free expanding phase is \citep{2001AA...380..309V,2003AA...404..939V,2014JHEAp...1...31T}
\begin{equation}
R_{\mathrm{PWN}}(t) = 0.84 \left( \frac{L_0 t}{E_0} \right)^{1/5} \left( \frac{10E_0}{3M_{\mathrm{ej}}} \right )^{1/2}t,
\label{eq:rpwn}
\end{equation}
where $E_0 =10^{51}\mathrm{erg}$ is the knetic energy of the supernova ejecta with a mass of $M_{\mathrm{ej}}=9.5M_{\odot}$ for the PWN in this paper. $\eta=0.03$ is adopt in \citet{2014JHEAp...1...31T} to model the Crab nebula, and we also use this fraction in this paper.

Leptons are continually injected into the PWN with a distribution of a power-law \citep{1984ApJ...283..710K,1984ApJ...278..630R}
\begin{equation}
Q(\gamma, t) = Q_0(t) \gamma^{-\alpha},
\label{eq:qgamma1}
\end{equation}
or a broken power-law \citep{1984ApJ...283..710K,1984ApJ...278..630R}
\begin{equation}
Q(\gamma, t) = Q_0(t) \left\{
  \begin{array}{cc}
    \left( \frac{\gamma}{\gamma_{\mathrm{b}}} \right)^{-\alpha_1} & \mathrm{if~} \gamma \leq \gamma_{\mathrm{b}}\;, \\
    \left( \frac{\gamma}{\gamma_{\mathrm{b}}} \right)^{-\alpha_2} & \mathrm{if~} \gamma_{\mathrm{b}} < \gamma \leq \gamma_{\mathrm{max}}\;.
    \label{eq:qgamma2}
   \end{array}
\right.
\end{equation}
$Q_0(t)$ is given by \citep{2012mnras.427.415m,2014JHEAp...1...31T}
\begin{equation}
(1-\eta)L(t) = \int \gamma m_{\mathrm{e}}c^2Q(\gamma, t) d\gamma.
\label{eq:q0}
\end{equation}
The maximum energy of the particles is determined according to the Larmor radius  $R_{\mathrm L}$ must be smaller than a fraction of the radius of the termination shock $R_{\mathrm s}$, i.e., $R_{\mathrm L}=\varepsilon R_{\mathrm s}$, where $\varepsilon$ is the fractional size of the shock \citep{2012mnras.427.415m}.  A range of $0.2\leq\varepsilon\leq1/3$ is usually adopted to derive the maximum energy of the particles for PWNe \citep{2012mnras.427.415m,2014JHEAp...1...31T}.
The maximum Lorentz factor can be derived from \citep{2012mnras.427.415m,2014JHEAp...1...31T}
\begin{equation}
\gamma_{\mathrm{max}}(t) = \frac{\varepsilon e \kappa}{m_{\mathrm{e}}c^2}\left ( \frac{\eta L(t)}{c} \right )^{1/2},
\label{eq:gmax}
\end{equation}
where $m_{\mathrm{e}}$, $e$ are the electron mass and charge.  For strong shocks, the magnetic compression ratio at the shock is about $3$, and we  fix $\kappa$ to 3 in this paper \citep{2012AA...539A..24H,2012mnras.427.415m}.

After the particles are injected into the PWN, they encounter adiabatic and radiation losses. Multiband nonthermal emission is produced via synchrotron and inverse Compton scattering off the background soft photons, which consist of the synchrotron emission, cosmic-microwave background (CMB), and IR/optical photon fields. Details of the formulae on the two radiative processes used in the calculation can be found in the appendixes of \citet{2012mnras.427.415m}.

\section{Results}
\label{sect:results}
The rotational period of the pulsar PSR J2021$+$3651 is $0.104\times10^{-3}\,\mathrm{s}$ \citep{2002ApJ...577L..19R}. With the first period derivative $\dot{P}= 9.6\times10^{-14}\,\mathrm{s\,s}^{-1}$ , the characteristic age is $\tau_{\mathrm c}=1.7\times10^4\,\mathrm{yr}$, and the current luminosity is $3.4\times10^{36}\mathrm{erg\,s}^{-1}$. The distance of the pulsar has been estimated to be $1.8^{+1.7}_{-1.4}\,\mathrm{kpc}$ \citep{2015ApJ...802...17K}, and we adopt $1.8\,\mathrm{kpc}$ in this paper.  The three interstellar photon fields involved in calculating the spectrum of the inverse Compton scattering are CMB with temperature $T_{\mathrm{cmb}}=2.7\,\mathrm{K}$ and energy density $U_{\mathrm{cmb}}=0.25\,\mathrm{eV\,cm}^{-3}$, IR background with $T_{\mathrm{IR}}=30\,\mathrm{K}$ and $U_{\mathrm{IR}}=0.3\,\mathrm{eV\,cm}^{-3}$ \citep{2017ApJ...841..104M}, and star light (SL) with $T_{\mathrm{s}}=5000\,\mathrm{K}$ and $U_{\mathrm{s}}=1\,\mathrm{eV\,cm}^{-3}$.

\begin{figure}
        \centering
        \includegraphics[width=0.45\textwidth]{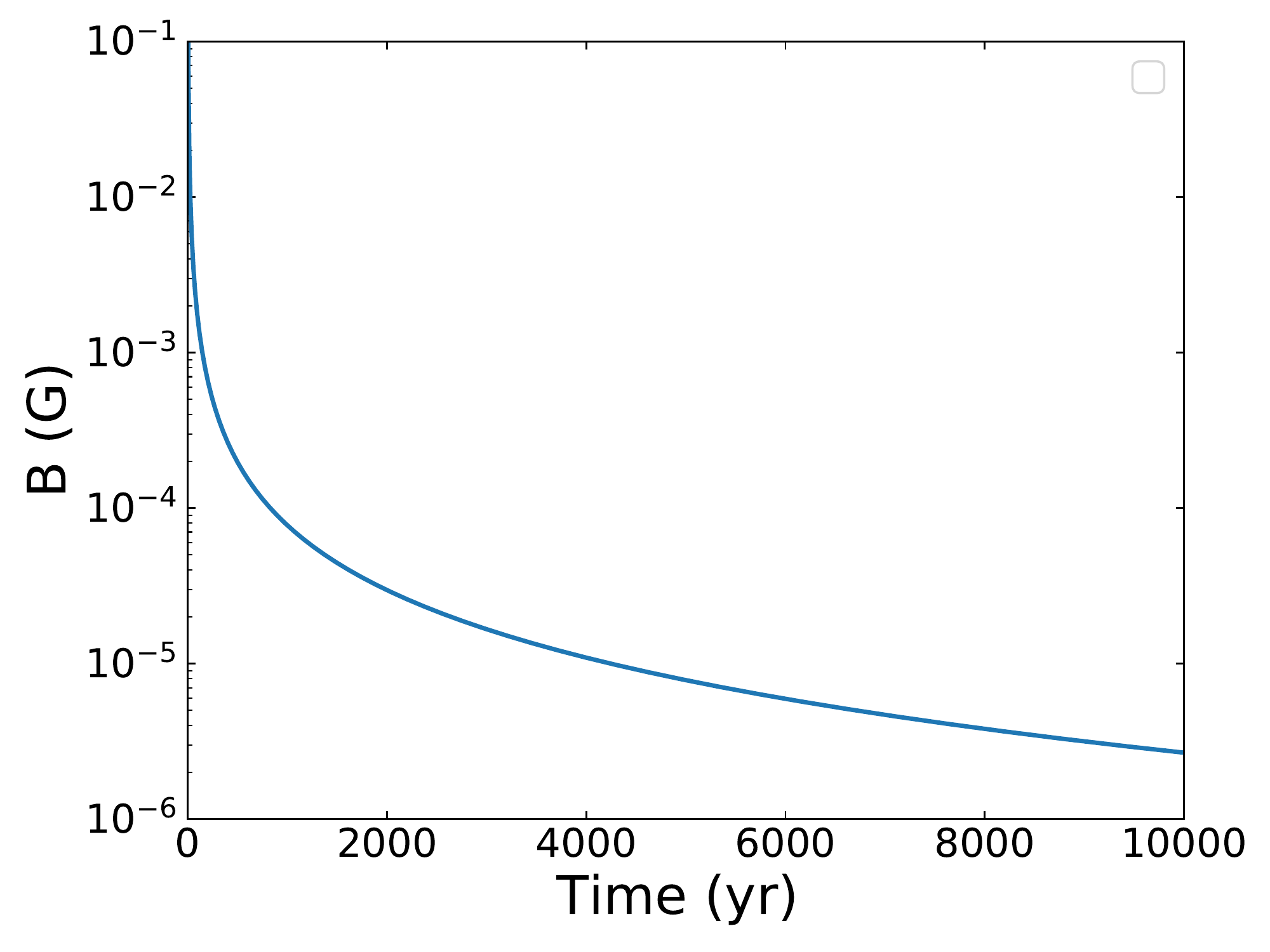}
        \caption{The evolution of the magnetic field strength in the PWN with $\eta=0.03$.}
        \label{fig:bt}
\end{figure}

\begin{figure}
        \centering
        \includegraphics[width=0.45\textwidth]{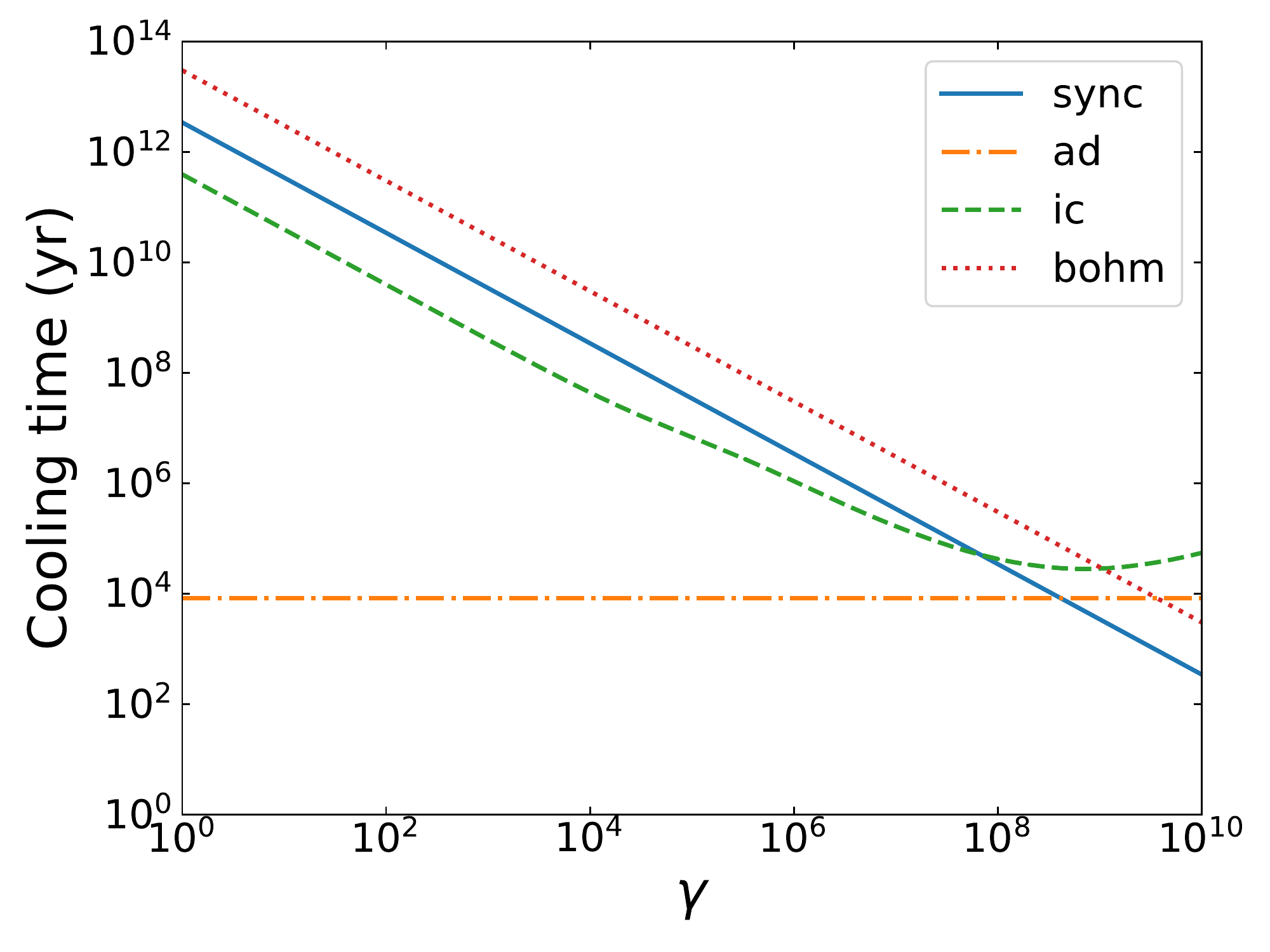}
        \caption{Cooling times for synchrotron radiation, adiabatic loss, inverse Compton scattering and the escape time for Bohm diffusion at $t_{\mathrm{age}}=10^4\,\mathrm{yr}$.}
        \label{fig:coolingtime}
\end{figure}

The age of PSR J2021$+$3651 is unknown, and we firstly adopt it to be $10^4\,\mathrm{yr}$. With this age, the initial spin-down time scale is $\tau_0 = 7.15\times10^3\,\mathrm{yr}$, and the initial luminosity is $L_0=1.95\times10^{37}\mathrm{erg\,s}^{-1}$.  \citet{2014JHEAp...1...31T} investigated the multiband radiative properties of some TeV-detected PWNe, the energies contained in the particles and the magnetic field do not satisfy equipartition. The magnetic energy ratio $\eta$ is usually $<0.1$ to reproduce the detected multiband fluxes for the PWNe.   As illustrated in Fig.\ref{fig:bt} and  {\bf Eq.\ref{eq:bt}}, the magnetic field decreases with time due to the expansion of the nebula.  With $\eta=0.03$, which is the same as the Crab nebula in \citet{2014JHEAp...1...31T}, the magnetic field strength in the nebula is $2.7\,\mu \mathrm{G}$ at a time of $10^4\mathrm{yr}$. Fig.\ref{fig:coolingtime} indicates the cooling times for the synchrotron radiation, the adiabatic loss, the inverse Compton scattering and the escape time due to the Bohm diffusion for the particles with different lorentz factor. For $\gamma<10^8$, the adiabatic loss is the dominate process to cool the particles. For $5\times10^8 < \gamma < 5\times10^9$, the synchrotron radiation becomes the most prominent process, and the particles with $\gamma\sim10^{10}$ escape quickly from the nebula via the Bohm diffusion.

\begin{figure}
        \centering
        \includegraphics[width=0.5\textwidth]{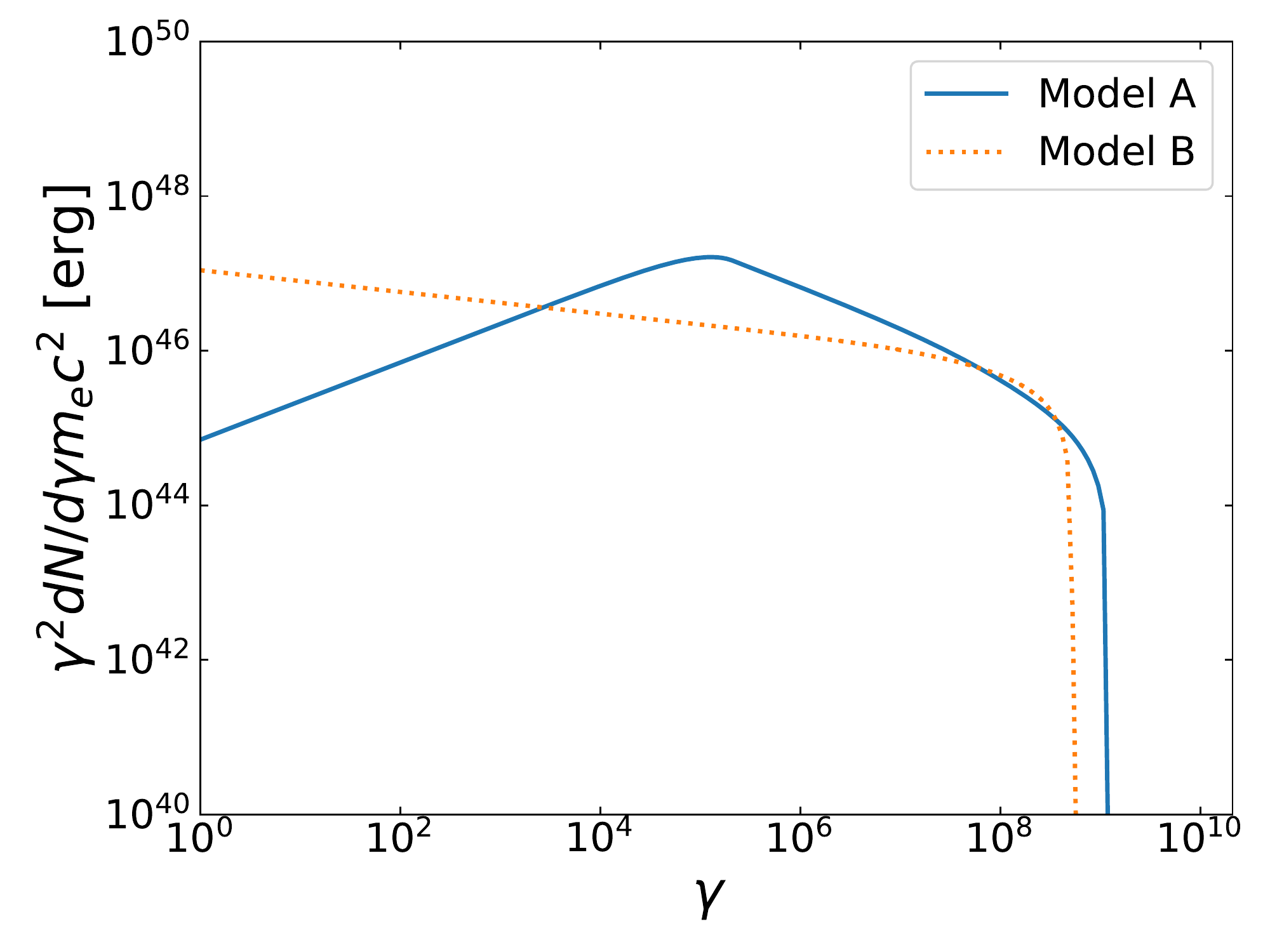}
        \caption{Particle distributions of the electrons/positrons in the nebula on  $\gamma$ for the Models A and B at $t_{\mathrm{age}}=10^4\,\mathrm{yr}$.}
        \label{fig:pardis}
\end{figure}

\begin{table*}
\caption{Summary of observations}
\label{Tab:obs}
\begin{center}
\small
\begin{tabular}{ccccc}
\hline\hline
Observatory     & $E^2dN/dE$ & energy (range) & source name (extension) & reference \\
                & $\mathrm{erg} \, \mathrm{cm}^{-2} \,\mathrm{s}^{-1}$ & MeV & & \\

\hline
VLA    &  $5.8\times10^{-12}$ & $9.8\times10^{-15}$ & ($\sim 100'^2$) & \citet{2009AA...507..241P} \\
\textit{Suzaku, XMM-Newton} & $7\times10^{-12}$ & (2 -- 10) $\times10^{-3}$ & ($\sim 30'\times12'$) & \citet{2017ApJ...841..104M} \\
VERITAS &   & (1 -- 30) $\times 10^6$  & VER J2019$+$368 ($\sim 20.4'\times7.8'$) & \citet{2014ApJ...788...78A} \\
HAWC &   & (1 -- 177) $\times 10^6$  & eHWC J2019$+$368 ($0.20\pm0.05^{\circ}$) & \citet{2020prl...124..021102A} \\
Milagro & $(5.6\pm1.7)\times10^{-12}$  &  $12\times10^6 $ & MGRO J2019+37 & \citet{2007ApJ...658L..33A} \\
                 & $(5.6\pm0.9)\times10^{-12}$ & $20\times10^6$  & MGRO J2019+37 ($1.1\pm0.5^{\circ}$)  & \citet{2007ApJ...664L..91A} \\
                 &  $(2.1\pm0.2)\times10^{-12}$  & $35\times10^6$  & MGRO J2019+37 & \citet{2009ApJ...700L.127A} \\
\hline
\end{tabular}
\end{center}
\end{table*}

\begin{figure}
        \centering
        \includegraphics[width=0.5\textwidth]{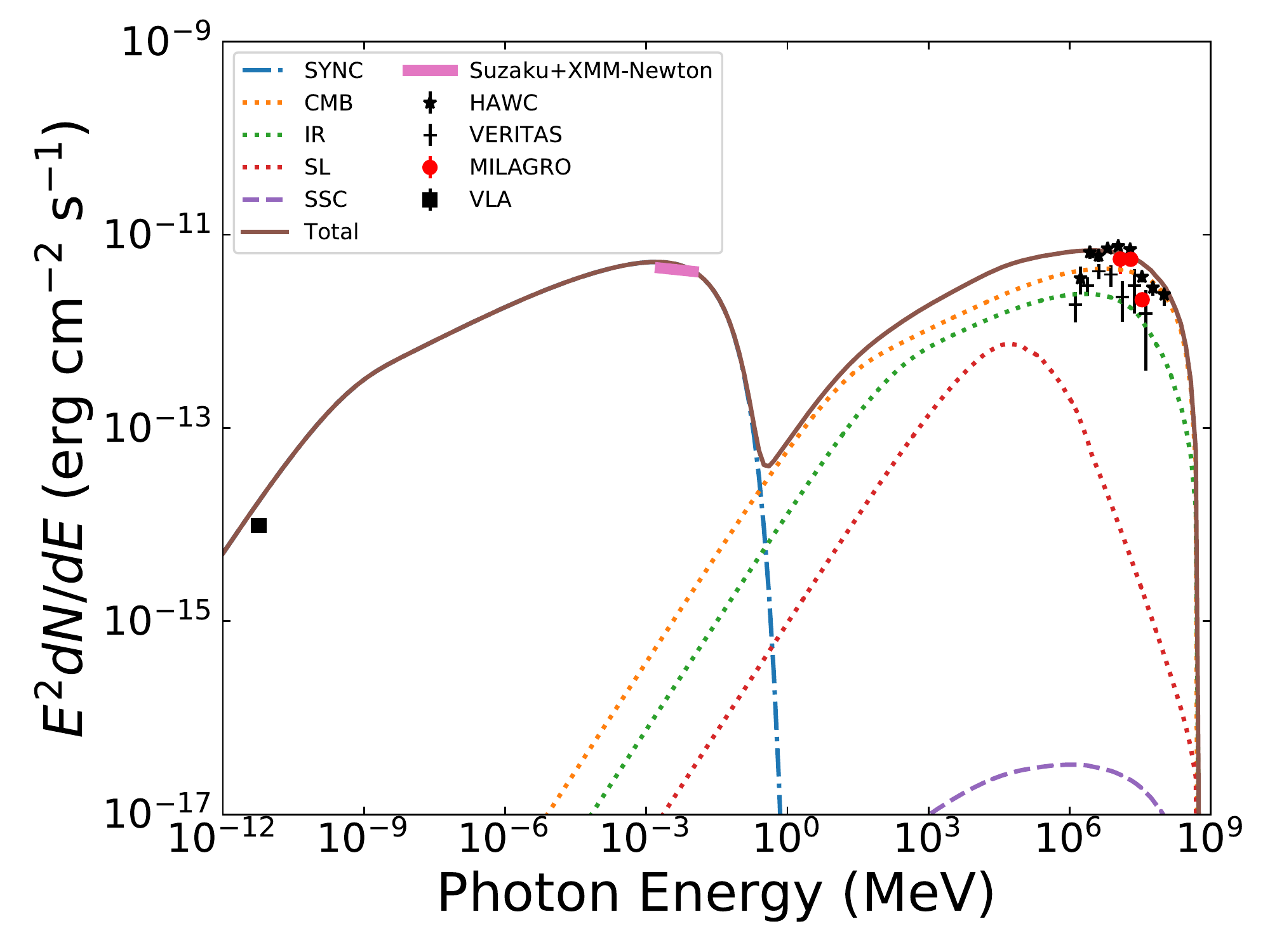}
        \caption{Spectral energy distribution from the model for the PWN G75.2$+$0.1 with $t_{\mathrm{age}}=10^4\,\mathrm{yr}$. The observed fluxes in the radio, X-ray and $\gamma$-ray bands are taken from \citet{2009AA...507..241P}, \citet{2017ApJ...841..104M}, \citet{2007ApJ...658L..33A,2007ApJ...664L..91A,2009ApJ...700L.127A}, \citet{2014ApJ...788...78A} and \citet{2020prl...124..021102A}. }
        \label{fig:j2019}
\end{figure}

With the assumption that the particles injected in G75.2$+$0.1 have a broken power-law distribution with $\alpha_1=1.5$, $\alpha_2=2.5$, $\gamma_b=2.0\times 10^5$ and $\varepsilon=1/3$  (Model A).  At $t_{\mathrm{age}}=10^4\,\mathrm{yr}$, as indicated in Fig.\ref{fig:pardis}, the particle spectrum has a break at $\gamma_b$ and a cutoff Lorentz factor $\sim 1.2\times 10^9$ ($\sim0.6$ PeV). The resulting photon spectra at $t_{\mathrm{age}}=10^4\,\mathrm{yr}$ from the synchrotron, the SSC and the inverse Compton scattering off the CMB, IR background and SL are illustrated in Fig.\ref{fig:j2019}.  The radio flux density detected with the Very Large Array (VLA) is $700$\,mJy at $1.4\,\mathrm{GHz}$ for the extend emission with an extension of $\sim 100'^2$ in the vicinity of the PWN G75.2$+$0.1 \citep{2009AA...507..241P}. The X-ray spectrum derived from the observations with {\it Suzaku } and {\it XMM-Newton} corresponds to a flux of $7\times10^{-12}\,\mathrm{erg}\,\mathrm{cm}^{-2}\,\mathrm{s}^{-1} $ with a photon index of $2.05$ in 2-10 keV for the overall PWN \citep{2017ApJ...841..104M}. The fluxes with  VERITAS for the source VER J2019$+$368 and Milagro for MGRO J2019+37 at 12, 20, 35 TeV are also shown in Fig.\ref{fig:j2019}. The spectrum of VER J2019+378 detected by VERITAS can be fitted with a power-law with an index of $\sim 1.75$ in  1 -- 10 TeV \citep{2014ApJ...788...78A} and the detected $\gamma$-ray flux is lower than that with  HAWC. The observational results are summarized in Table \ref{Tab:obs}.  The synchrotron spectrum has a break at $\sim 10^{-9}\mathrm{MeV}$ due to the change of the index of the particle spectrum at $\gamma_b$, and the higher energy cutoff at several keV results from the energy losses. In the TeV band, the $\gamma$-rays  are mainly produced via the inverse Compton scattering off the CMB and IR photons, and the SSC process is negligible in producing the $\gamma$-rays with such a low magnetic field strength of $2.7\,\mu \mathrm{G}$.  The resulting spectrum from the inverse Compton scattering is consistent with those obtained by HAWC with energies above 3 TeV, but at lower energies the $\gamma$-ray flux from the model is significantly higher than  those by either HAWC or  VERITAS.

\begin{figure}
        \centering
        \includegraphics[width=0.5\textwidth]{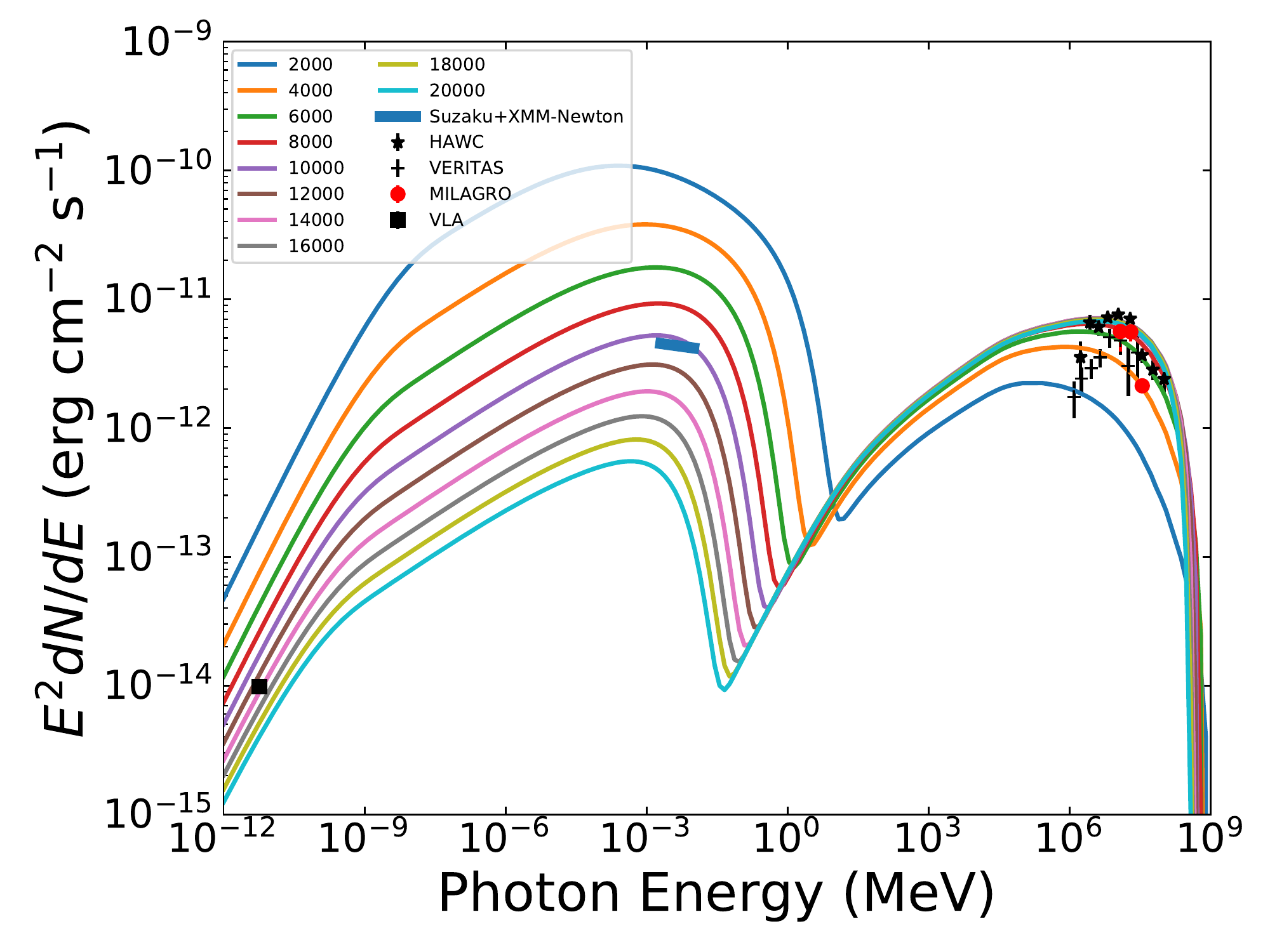}
        \caption{The spectral energy distributions at different times from the model for the PWN G75.2$+$0.1 with $t_{\mathrm{age}}=10^4\,\mathrm{yr}$. The others are the same as Fig.\ref{fig:j2019}. }
        \label{fig:j2019dw}
\end{figure}

With an age of $10^4\,\mathrm{yr}$, the spectral energy distributions at different times are indicated in Fig.\ref{fig:j2019dw}. The synchrotron radiation diminishes significantly with time because the magnetic field decreases due to the expansion of the nebula. In the $\gamma$-ray band above 10 MeV, the spectrum from the inverse Compton scattering after a time of $\sim 10^4\,\mathrm{yr}$ saturates due to the balance of the injection of the high-energy particles and the processes of the energy losses.
\begin{figure}
        \centering
        \includegraphics[width=0.5\textwidth]{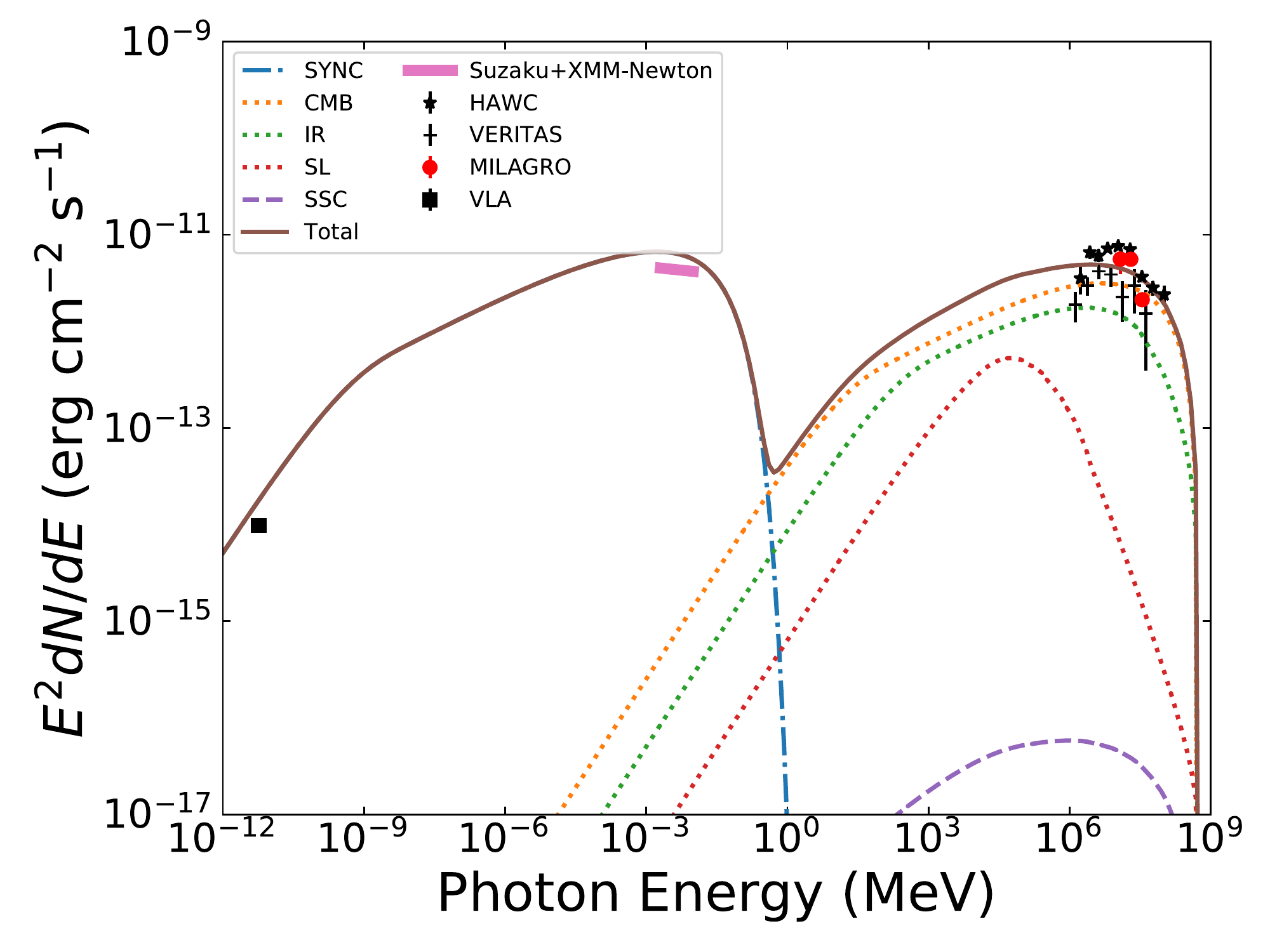}
        \caption{Spectral energy distribution from the model for G75.2$+$0.1 with $t_{\mathrm{age}}=8\times10^3\,\mathrm{yr}$. The others are the same as Fig.\ref{fig:j2019}. }
        \label{fig:j20198000}
\end{figure}
\begin{figure}
        \centering
        \includegraphics[width=0.5\textwidth]{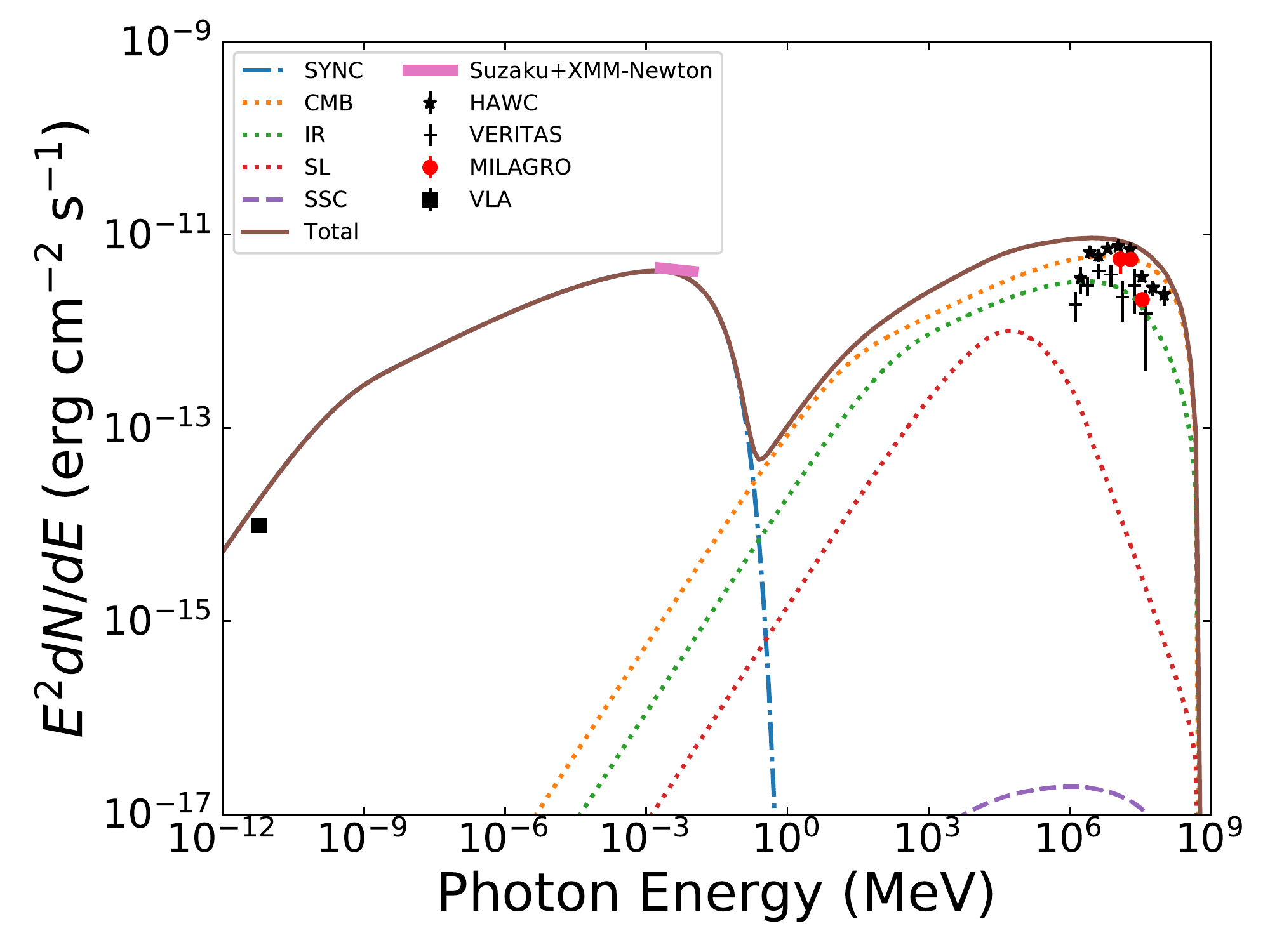}
        \caption{Spectral energy distribution from the model for G75.2$+$0.1 with $t_{\mathrm{age}}=1.2\times10^4\,\mathrm{yr}$. The others are the same as Fig.\ref{fig:j2019}. }
        \label{fig:j201912000}
\end{figure}

The effect of the age on the resulting multiband spectral energy distribution of the PWN is illustrated in Fig.\ref{fig:j20198000} and Fig.\ref{fig:j201912000}. With an age of $8\times10^3\,\mathrm{yr}$, the initial luminosity is $L_0=1.2\times10^{37}\mathrm{erg\,s}^{-1}$, and less energy is injected in the nebula compared with $t_{\mathrm{age}}=10^4\,\mathrm{yr}$. As illustrated in Fig.\ref{fig:j20198000}, the resulting flux in the band of 1 -- 30 $\mathrm{TeV}$ is lower than that obtained with HAWC. Whereas for $t_{\mathrm{age}}=1.2\times10^4\,\mathrm{yr}$, a higher initial luminosity $L_0=3.8\times10^{37}\mathrm{erg\,s}^{-1}$ is adopted, and the resulting $\gamma$-ray flux is relatively higher than the HAWC result (see Fig.\ref{fig:j201912000}). Therefore, with the energy densities of the background radiation field and the broken power-law spectrum for the high-energy particles used in the model, an age of $10^4\,\mathrm{yr}$ is preferred to reproduce the detected TeV $\gamma$-ray flux with HAWC.

\begin{figure}
        \centering
        \includegraphics[width=0.5\textwidth]{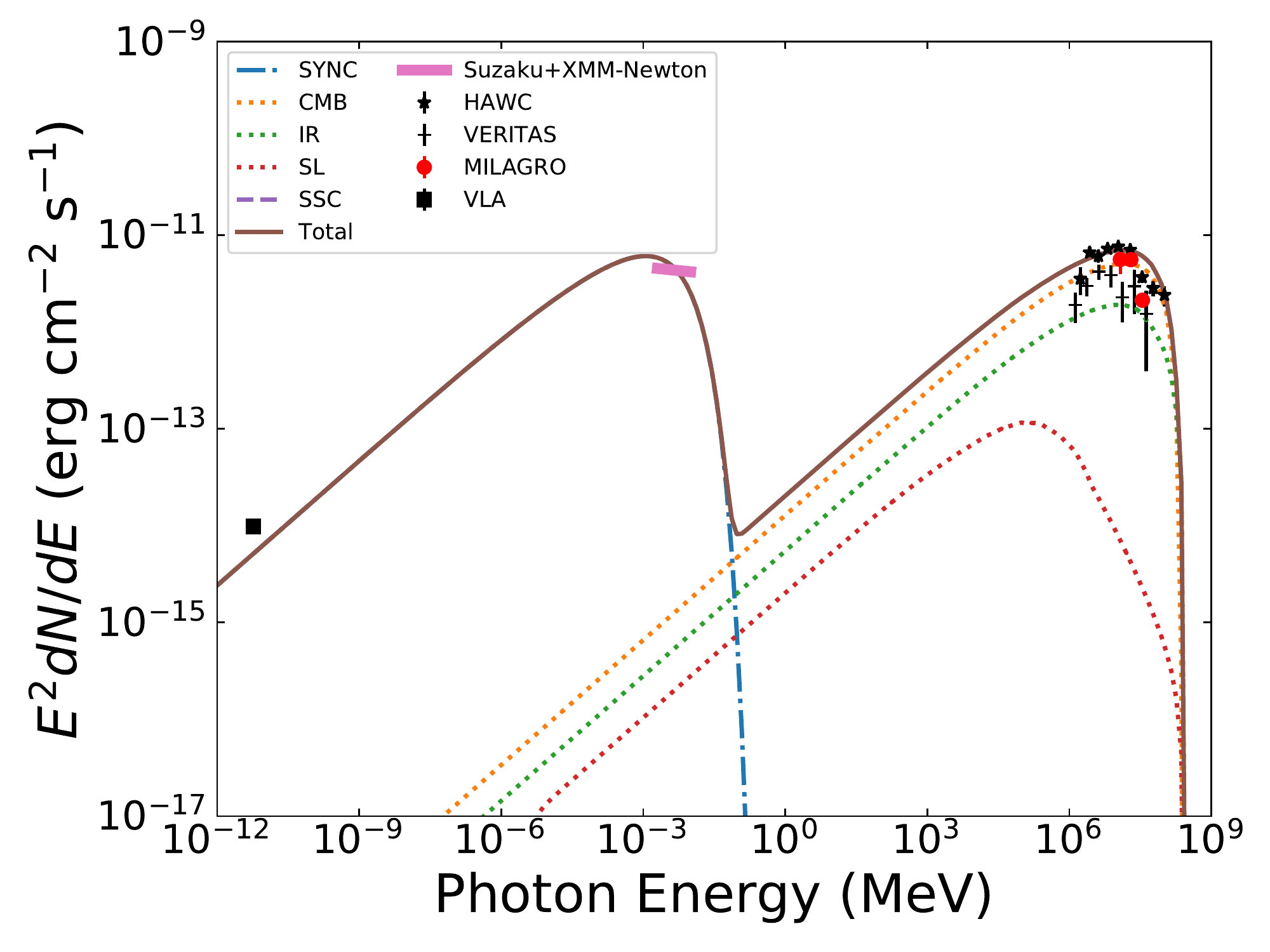}
        \caption{Spectral energy distribution from the model for G75.2$+$0.1 with $t_{\mathrm{age}}=1.0\times10^4\,\mathrm{yr}$ and a power-law spectrum of the injected particles ($\alpha=2.14$). The others are the same as Fig.\ref{fig:j2019}. }
        \label{fig:j2019216}
\end{figure}

\begin{figure}
        \centering
        \includegraphics[width=0.5\textwidth]{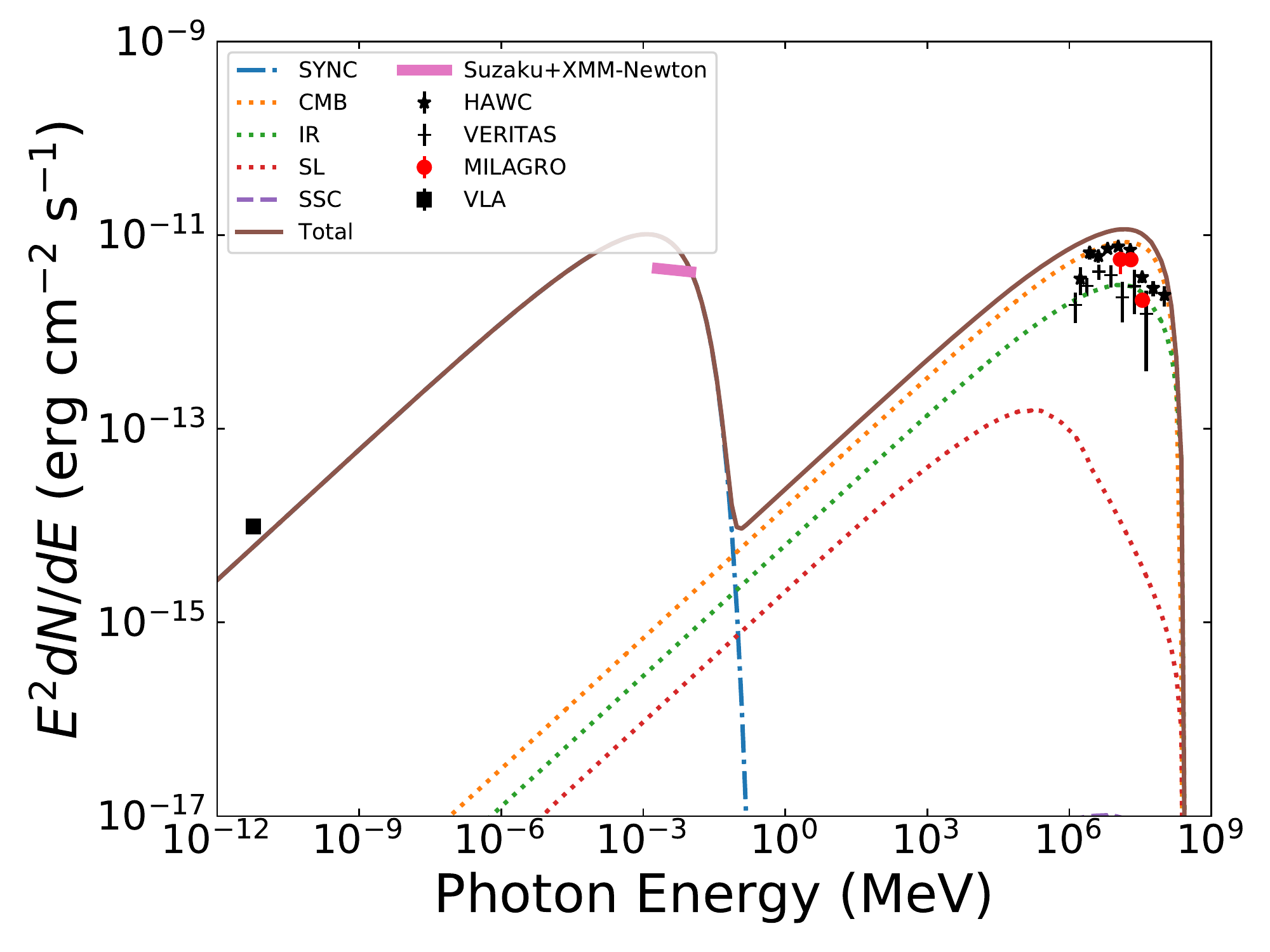}
        \caption{Spectral energy distribution from the model for G75.2$+$0.1 with a power-law spectrum of the injected particles ($\alpha=2.1$). The others are the same as Fig.\ref{fig:j2019216}. }
        \label{fig:j2019216ed2}
\end{figure}

 Fig.\ref{fig:j2019216} indicates the resulting multiband spectrum assuming the particle injected into the nebula has a distribution of a power-law with $\alpha=2.14$ and $t_{\mathrm{age}}=10^4\,\mathrm{yr}$ (Model B).  The maximum energy of the particles in the nebula and that of the most energetic $\gamma$-rays depend sensitively on $\varepsilon$, which is constrained to $0.15$ to reproduce the spectrum in the TeV $\gamma$-rays measured with HAWC.  The leptons have a power-law distribution on the lorentz factor with a cutoff energy of $\sim0.4$ PeV (Fig.\ref{fig:pardis}), and the break of the resulting spectrum for the synchrotron at $\sim 10^{-9}\,\mathrm{MeV}$ disappears. The resulting flux in the band 1 -- 10 TeV is significantly lower than that with the broken power-law distribution for the injected particles in the Model A. The flux in 3 -- 20 TeV is consistent well with that detected by HAWC, and the slope of the spectrum in 1 -- 10 TeV obtained by VERITAS can be better reproduced than the Model A.  The multi-band spectral energy distribution depends sensitively on the index of the power-law distribution for the particles. Fig.\ref{fig:j2019216ed2} shows the spectral energy distribution for the nebula with $\alpha=2.1$ and the other parameters same as the Model B. With such a harder spectrum, the fluxes both in the X-rays and in the $\gamma$-rays are higher than the detected ones.

\section{summary and discussion}
\label{sect:discon}
We study the multiband nonthermal emission of the PWN G75.2$+$0.1 based on the time-dependent model, and investigate whether the nebula has the ability to produce the TeV $\gamma$-rays of eHWC J2019+368. Assuming the electrons/positrons injected into the nebula has the broken power-law distribution, the resulting spectral energy distribution of the PWN is generally consistent with the fluxes detected by  VLA, {\it Suzaku} + {\it XMM-Newton} and HAWC except at $1.7$ TeV. With the power-law spectrum for the injected particles in the Model B, the $\gamma$-ray flux at 1.7 TeV detected by HAWC and the slope of the spectrum in the band 1 -- 10 TeV obtained by VERITAS can be better reproduced than the Model A. Our studies indicate that the PWN G75.2$+$0.1 powered by the energetic pulsar PSR J2021$+$3651 is capable of producing the $\gamma$-ray flux of eHWC J2019+368 with reasonable parameters, and the energetic electrons/positrons in the nebula have energies up to $\sim 0.4$ PeV.

The $\gamma$-ray flux resulting from the inverse Compton scattering depends on the age of the pulsar, because more energy can be injected into the high-energy electrons/positrons in the nebula with an older age. Since the injection of the high-energy particles is balanced by the energy losses after a certain time, the TeV $\gamma$-rays resulting from the inverse Compton scattering saturates.  With $\eta=0.03$, an age of $10^4\,\mathrm{yr}$ is derived in this paper to reproduce a $\gamma$-ray flux consistent with the measurements in the TeV band. However, the preferred age depends sensitively on the distance of the pulsar.  A larger distance implies an older age of the pulsar to reproduce the detected $\gamma$-ray flux in the TeV band.

The Crab nebula is also a $\gamma$-ray source emitting photons with energies past 100 TeV. Assuming the particles injected into the nebula have a spectrum of a broken power-law with $\alpha_1=1.5$, $\alpha_2=2.5$ and $\gamma_b = 7\times10^5$, the multiband spectrum consists well with the observed fluxes from radio to $\gamma$-rays \citep{2014JHEAp...1...31T}. The two indexes are the same as those used in the Model A of this paper for the PWN powered by PSR J2021$+$3651. The initial luminosity of the Crab pulsar is $\sim3.1\times10^{39}\mathrm{erg\,s}^{-1}$ \citep{2014JHEAp...1...31T}, which is about 16 times higher than that PSR J2021$+$3651 in this paper. However, the inferred age for PSR J2021$+$3651 is about 10 times older than the Crab pulsar, and the spin-down energy has been injected into the nebula for a longer time compared with the Crab nebula. As a result, G75.2$+$0.1 has the comparable ability to produce $\gamma$-rays as the Crab nebula.

Some of the detected TeV $\gamma$-rays of eHWC J2019+368  may be originated from the other sources in the Cygnus region besides G75.2$+$0.1, such as the particles accelerated from the ensemble of OB associations \citep{2014ChPhc.38.085001H} and the H{\sc ii} region
Sh 2-104 \citep{2014ApJ...788...78A}. Further observations with the Large High Altitude Air Shower Observatory (LHAASO) will give more insights into the origin the high-energy particles producing the TeV $\gamma$-rays of eHWC J2019+368 \citep{2019arXiv190502773B}.

\section*{Acknowledgements}
JF is partially supported by the National Key R\&D Program of China under grant No.2018YFA0404204, the
Natural Science Foundation of China (NSFC) through grants 11873042 and 11563009,
the Yunnan Applied Basic Research
Projects under (2018FY001(-003)), the Candidate Talents Training Fund of Yunnan Province (2017HB003)
and the Program for Excellent Young Talents, Yunnan University (WX069051, 2017YDYQ01).

\setlength{\bibhang}{2.0em}
\setlength\labelwidth{0.0em}
\bibliography{PWN100tev2019}

\begin{thebibliography}{}
\providecommand{\href}[2]{#2}
  \providecommand{\doi}[1]{\href{http://dx.doi.org/#1}{doi:#1}}
  \providecommand{\eprint}[1]{\href{http://arxiv.org/abs/#1}{arXiv:#1}}

\bibitem[\protect\citeauthoryear{{Abdo} et~al.,}{{Abdo}
  et~al.}{2009}]{2009ApJ...700..1059A}
Abdo, A. A., Ackermann, M., Ajello, M., et al. 2009b, ApJ, 700, 1059

\bibitem[\protect\citeauthoryear{{Abdo} et~al.,}{{Abdo}
  et~al.}{2009a}]{2009ApJ...700L.127A}
Abdo, A. A., Allen, B. T., Aune, T., et al. 2009a, ApJ, 700, L127

\bibitem[\protect\citeauthoryear{{Abdo} et~al.,}{{Abdo}
  et~al.}{2012}]{2012ApJ...753..159A}
Abdo, A. A., Allen, B. T., Aune, T., et al. 2012, ApJ, 753, 159

\bibitem[\protect\citeauthoryear{{Abdo} et~al.,}{{Abdo}
  et~al.}{2007a}]{2007ApJ...658L..33A}
Abdo, A. A., Allen, B., Berley, D., et al. 2007a, \apj, 658, L33

\bibitem[\protect\citeauthoryear{{Abdo} et~al.,}{{Abdo}
  et~al.}{2007b}]{2007ApJ...664L..91A}
Abdo, A. A., Allen, B. T., Berley, D., et al. 2007b, ApJ, 664, L91

\bibitem[\protect\citeauthoryear{{Abeysekara} et~al.,}{{Abeysekara}
  et~al.}{2018}]{2018ApJ...861..134A}
Abeysekara, A. U., et al., 2018, \apj, 2018, 861, 134

\bibitem[\protect\citeauthoryear{{Abeysekara} et~al.,}{{Abeysekara}
  et~al.}{2020}]{2020prl...124..021102A}
Abeysekara, A. U., et al., 2020, \prl, 124, 021102

\bibitem[\protect\citeauthoryear{{Aliu} et~al.,}{{Aliu}
  et~al.}{2014}]{2014ApJ...788...78A}
Aliu, E., Aune, T., Behera, B., et al. 2014, ApJ, 788, 78

\bibitem[\protect\citeauthoryear{{Amato},}{{Amato}
  }{2020}]{2020arXiv200104442A}
Amato, E. 2020, arXiv:2001.04442
\bibitem[\protect\citeauthoryear{{Bai} et~al.,}{{Bai}
  et~al.}{2019}]{2019arXiv190502773B}
Bai, X., Bi, B. Y., Bi, X. J., et al. 2019, arXiv:1905.02773

\bibitem[\protect\citeauthoryear{{Bartoli} et~al.,}{{Bartoli}
  et~al.}{2012}]{2012ApJ...745L..22B}
Bartoli, B., Bernardini, P., Bi, X. J., et al. 2012, \apj, 745, L22

\bibitem[\protect\citeauthoryear{{Bi} et~al.,}{{Bi}
  et~al.}{2009}]{2009ApJ...695..883B}
Bi, X.-J., Chen, T.-L., Wang, Y., \& Yuan, Q. 2009, ApJ, 695, 883

\bibitem[\protect\citeauthoryear{{Crumley} et~al.,}{{Crumley}
  et~al.}{2019}]{2019MNRAS.485.5105C}
Crumley, P.,  Caprioli, D.,  Markoff, S.,  Spitkovsky, A., 2009, \mnras, 485, 5105

\bibitem[\protect\citeauthoryear{{Fang} et~al.,}{{Fang}
  et~al.}{2018}]{2018MNRAS.474.2544F}
Fang, J., Yu, H., Zhang, L., 2018, \mnras, 474, 2544

\bibitem[\protect\citeauthoryear{{Fang \& Zhang}}{{Fang \& Zhang}
  }{2010}]{2010AA...515A..20F}
Fang, J., Zhang, L. 2010, \aap, 525, A20


\bibitem[\protect\citeauthoryear{{Gaensler \& Slane}}{{Gaensler \& Slane}
  }{2006}]{2006araa.44.17G}
Gaensler, B. M., Slane, P. O., 2006, Annu. Rev. Astron. Astrophys. 44, 17

\bibitem[\protect\citeauthoryear{{Hessels} et~al.,}{{Hessels}
  et~al.}{2004}]{2004ApJ...612..389H}
Hessels, J. W. T., Roberts, M. S. E., Ransom, S. M., et al. 2004, ApJ, 612, 389

\bibitem[\protect\citeauthoryear{{Holler} et~al.,}{{Holler}
  et~al.}{2012}]{2012AA...539A..24H}
Holler, M., et al., 2012, A\&A, 539. A24.

\bibitem[\protect\citeauthoryear{{Hou} et~al.,}{{Hou}
  et~al.}{2014}]{2014ChPhc.38.085001H}
Hou, C., Chen, S.-Z., Yuan, Q., et al. 2014, ChPhC, 38, 085001

\bibitem[\protect\citeauthoryear{{Kennel \& Coroniti}}{{Kennel \& Coroniti}
  }{1984}]{1984ApJ...283..710K}
Kennel, C. F., \& Coroniti, F. V. 1984, ApJ, 283, 710


\bibitem[\protect\citeauthoryear{{Kirichenko} et~al.,}{{Kirichenko}
  et~al.}{2015}]{2015ApJ...802...17K}
Kirichenko, A., Danilenko, A., Shternin, P., et al. 2015, \apj, 802, 17


\bibitem[\protect\citeauthoryear{{Mart\'{\i}n} et~al.,}{{Mart\'{\i}n}
  et~al.}{2012}]{2012mnras.427.415m}
Mart\'{\i}n, J., Torres, D. F., Rea, N., 2012, MNRAS, 427,415.

\bibitem[\protect\citeauthoryear{{Mizuno} et~al.,}{{Mizuno}
  et~al.}{2017}]{2017ApJ...841..104M}
Mizuno, T.,  Tanaka, N., Takahashi, H., et al. 2017, \apj, 841, 104

\bibitem[\protect\citeauthoryear{{Paredes} et~al.,}{{Paredes}
  et~al.}{2009}]{2009AA...507..241P}
Paredes, J. M., Mart\'{\i}, J., Ishwara-Chandra, C. H., et al. 2009, \aap, 507, 241

\bibitem[\protect\citeauthoryear{{Reynolds \& Chevalier}}{{Reynolds \& Chevalier}
  }{1984}]{1984ApJ...278..630R}
Reynolds, S. P., \& Chevalier, R. A. 1984, ApJ, 278, 630

\bibitem[\protect\citeauthoryear{{Roberts} et~al.,}{{Roberts}
  et~al.}{2002}]{2002ApJ...577L..19R}
Roberts, M. S. E., Hessels, J. W. T., Ransom, S. M., et al. 2002, ApJL,
577, L19

\bibitem[\protect\citeauthoryear{{Spitkovsky}}{{Spitkovsky}}{2008}]{2008ApJ...682L...5S}
Spitkovsky, A. 2008, \apj, 682, L5


\bibitem[\protect\citeauthoryear{{Tanaka \& Takahara}}{{Tanaka \& Takahara}
  }{2010}]{2010apj.715.1248T}
Tanaka, S. J., Takahara, F., 2010, \apj, 715, 1248

\bibitem[\protect\citeauthoryear{{Torres} et~al.,}{{Torres}
  et~al.}{2014}]{2014JHEAp...1...31T}
Torres, D. F., Cillis, A., Mart\'{\i}n J., et al. 2014, J. High Energy
Astrophys, 1, 31

\bibitem[\protect\citeauthoryear{{van der Swaluw} et~al.,}{{van der Swaluw}
  et~al.}{2001}]{2001AA...380..309V}
van der Swaluw E., Achterberg A., Gallant Y. A., et al. 2001, A\&A,
380, 309
\bibitem[\protect\citeauthoryear{{van der Swaluw} et~al.,}{{van der Swaluw}  et~al.
  }{2003}]{2003AA...404..939V}
van der Swaluw, E. 2003, \aap, 404, 309

\bibitem[\protect\citeauthoryear{{Van Etten} et~al.,}{{Van Etten}  et~al.
  }{2008}]{2008ApJ...680.1417V}
Van Etten, A., Romani, R., W., Ng, C. -Y. 2008, \apj, 680, 1417

\bibitem[\protect\citeauthoryear{{Zhang} et~al.,}{{Zhang}
  et~al.}{2008}]{2008ApJ...676..1210Z}
Zhang L., Chen S. B., Fang J., 2008, ApJ, 676, 1210
\end{thebibliography}

\end{document}